\documentclass{osa-article}

\journal{oe}

\usepackage{newtxtext,newtxmath}
\usepackage{gensymb}
\usepackage{graphicx}	
\usepackage{amsmath}	
\usepackage{caption}

\begin{document}

\title{Adaptive hyperspectral imaging using structured illumination in a spatial light modulator-based interferometer}

\author{Amar Deo Chandra,\authormark{1} Mintu Karmakar,\authormark{1}, Dibyendu Nandy,\authormark{1,2} and Ayan Banerjee\authormark{1,2,*}}

\address{\authormark{1}Center of Excellence in Space Sciences India, Indian Institute of Science Education and Research Kolkata, Mohanpur 741246, West Bengal, India\\
\authormark{2}Department of Physical Sciences, Indian Institute of Science Education and Research Kolkata, Mohanpur 741246, West Bengal, India\\}

\email{\authormark{*}ayan@iiserkol.ac.in} 



\begin{abstract}
We develop a novel hyperspectral imaging system using structured illumination in an SLM-based Michelson interferometer. In our design, we use a reflective SLM as a mirror in one of the arms of a Michelson interferometer, and scan the interferometer by varying the phase across the SLM display. For achieving the latter, we apply a  checkerboard phase mask on the SLM display where the gray value varies between 0-255, thereby imparting a dynamic phase of up to 262$\degree$ to the incident light beam. We couple a supercontinuum source into the interferometer in order to mimic an astronomical object such as the Sun, and choose a central wavelength of 637.4 nm akin to the strong emission line of Fe X present in the solar spectrum. We use a bandwidth of 30 nm, and extract fringes corresponding to a spectral resolution of 3.8 nm which is limited by the reflectivity of the SLM.  We also demonstrate a maximum wavelength tunability of $\sim$8 nm by varying the phase over the phase mask with a spectral sampling of around 0.03 nm between intermediate fringes. The checkerboard phase mask can be adapted close to real time on time-scales of a few tens of milliseconds to obtain spectral information for other near-contiguous wavelengths. The compactness, potential low cost, low power requirements, real-time tunability and lack of moving mechanical parts in the setup implies that it can have very useful applications in settings which require near real-time, multi-wavelength spectroscopic applications, and is especially relevant in space astronomy.
\end{abstract}

\section{Introduction}
Hyperspectral imaging (HSI) entails acquiring spectral information in contiguous wavelength bands and spatial information of a given scene or sample. It mitigates a major issue in real-time spectroscopy - which is that of simultaneously obtaining high dynamic range and precision -  and as a result has applications in very diverse areas. Some of these include  remote sensing \cite{goetz1985imaging, sellar2005classification, ben2013hyperspectral},  coastal ocean imaging \cite{davis2002ocean}, camouflaged target detection and security applications \cite{manolakis2003, khan2018modern}, non-destructive testing of biological samples \cite{manley2014near}, 
non-contact investigation of forensic samples \cite{edelman2012hyperspectral, schuler2012preliminary} and non-invasive investigations of old heritage assets and archaeological samples \cite{fischer2006multispectral, liang2012advances, snijders2016using, cucci2016reflectance, odegaard2018underwater}. Recently they have also been employed in  aeronomy and astronomy research \cite{hagen2013review,mohanakrishna2018lenslet}, and some HSI instruments have actually been flown aboard space missions including the High-Resolution Imaging Spectrometer (HIRIS) \cite{goetz1991high}, Compact High-Resolution Imaging Spectrometer (CHRIS) \cite{barnsley2004proba}, Hyper-Spectral Imager (HySI) \cite{goswami2009chandrayaan}, Compact Reconnaissance Imaging Spectrometer for Mars (CRISM) \cite{murchie2007compact}. In addition, some of the HSI instruments have also been slated for future deployment, such as the Mapping Imaging Spectrometer for Europa (MISE) \cite{blaney2015mapping} and the Ultra-Compact Imaging Spectrometer Moon (UCIS-Moon) \cite{haag2020ultra} instruments. 

Some of the most commonly used instruments for hyperspectral imaging are whisk-broom imagers \cite{vane1993airborne}, pushbroom scanning imaging systems \cite{ mouroulis2000pushbroom, pearlman2003hyperion}, grating-based imaging spectrometers \cite{sigernes2000multipurpose}, snapshot imaging \cite{weitzel19963d} and more recently, electronically tunable filters \cite{gat2000imaging}. The advantages associated with using electronically tunable filters are compactness and the absence of mechanically moving parts. These filters use birefringent materials whose refractive index and concomitant spectral response can be modified in the presence of external stimuli such as electric voltage, light or temperature \cite{fuh2014optical}. 
Liquid crystal tunable filters (LCTFs) \cite{cenedese2006vegetation} and liquid crystal Fabry-Perot \cite{patel1990electrically, patel1991electrically, patel1992electrically, daly2000tunable} interferometers use liquid crystals whose refractive index can be tuned using external electric fields while acousto-optic tunable filters (AOTFs) use piezoelectric transducers bonded with a birefringent crystal \cite{harris1969acousto, dekemper2012tunable}. However, electronically tunable filters are sometimes challenged in achieving large tunability and precision simultaneously. 

In this paper, we report a novel method of adaptive hyperspectral imaging using structured illumination in a Spatial Light Modulator (SLM)-based interferometric configuration. The reflective SLM is placed in one arm of a Michelson interferometer and a checkerboard phase mask of gray level patterns is displayed on the SLM. An aperture mask generates a spatially multiplexed beam which is incident on the interferometer. We leverage the electrically tunable refractive index of the liquid crystal (LC)  structure of SLM to impart differential phase delay to each beamlet, obtaining near-contiguous spectral information for each beamlet in a single shot. Furthermore, the checkerboard pattern displayed on the SLM can be updated on short time-scales of about 18 ms (limited by the refresh rate of our SLM, can be even lesser for SLMs having faster tunability) recording another set of spectral information in another shot depending on requirements. The SLM needs to be phase calibrated \cite{chandra2020rapid} only one time before usage. The simplicity, compactness, potential low cost, low power requirements, absence of moving parts, and almost real-time tunability of the device can have very useful applications in settings which may require near real-time, multi-wavelength spectroscopic applications, that are galore in space astronomy. In addition, the phase mask is adaptive and can be updated in a few tens of ms to record another set of $\mathrm{N\times N}$ spectral information depending on requirements.  This method would therefore be very useful in settings wherein the optical emission and possibly concomitant spectral signature of the target varies quickly on time-scales of a few ms to few tens of ms, which renders obtaining the full spectral information using sequential scanning of an interferometer extremely challenging. A few applications in astronomical settings where our technique would be efficacious would include the analysis of optical emission from millisecond pulsars \cite{cocke1969discovery,wallace1977detection,middleditch1985optical,ambrosino2017optical}, and accretion discs around compact objects such as neutron stars and black holes \cite{motch1982discovery, beskin1994optical,bartolini1994fast}, solar dynamic activity, and optical follow-up of transient events such as $\gamma$-ray bursts \cite{kawai2006optical}.

\section{Methods and Results}
\subsection{Experimental setup}

\begin{figure}
\centering\includegraphics[width=\linewidth]{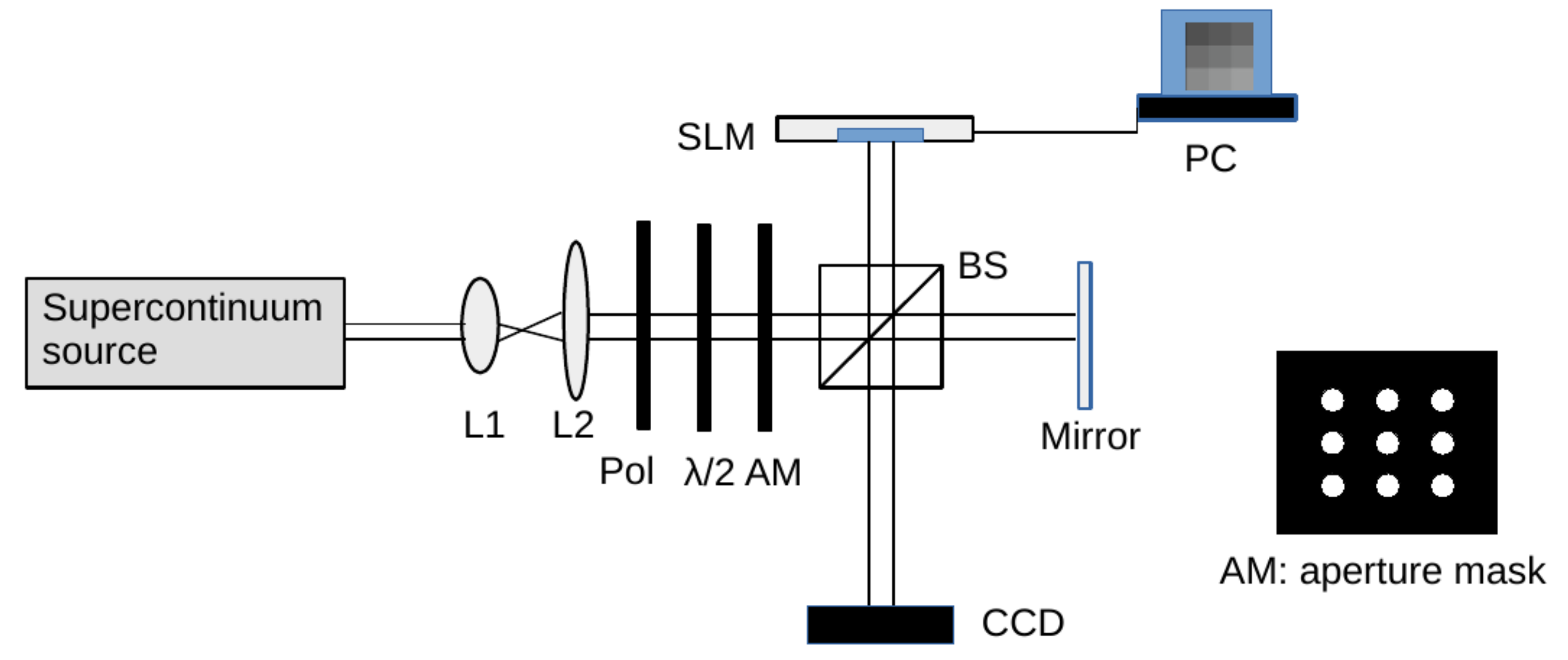}
\caption{Schematic diagram showing incident structured illumination in an SLM-based Michelson interferometer. Key: L1, L2: Lenses; Pol: Polarizer; $\lambda$/2: Half wave-plate; AM: Aperture mask having $\mathrm{3\times 3}$ holes; BS: Beam splitter. The plane mirror is mounted on a translation stage. For measuring the spectra from the Michelson interferometer, the CCD is replaced by a fiber spectrometer whose tip size is small enough to sample a single fringe pattern at a time.}
\label{fig1}
\end{figure}

\begin{figure}
\centering\includegraphics[width=0.5\linewidth]{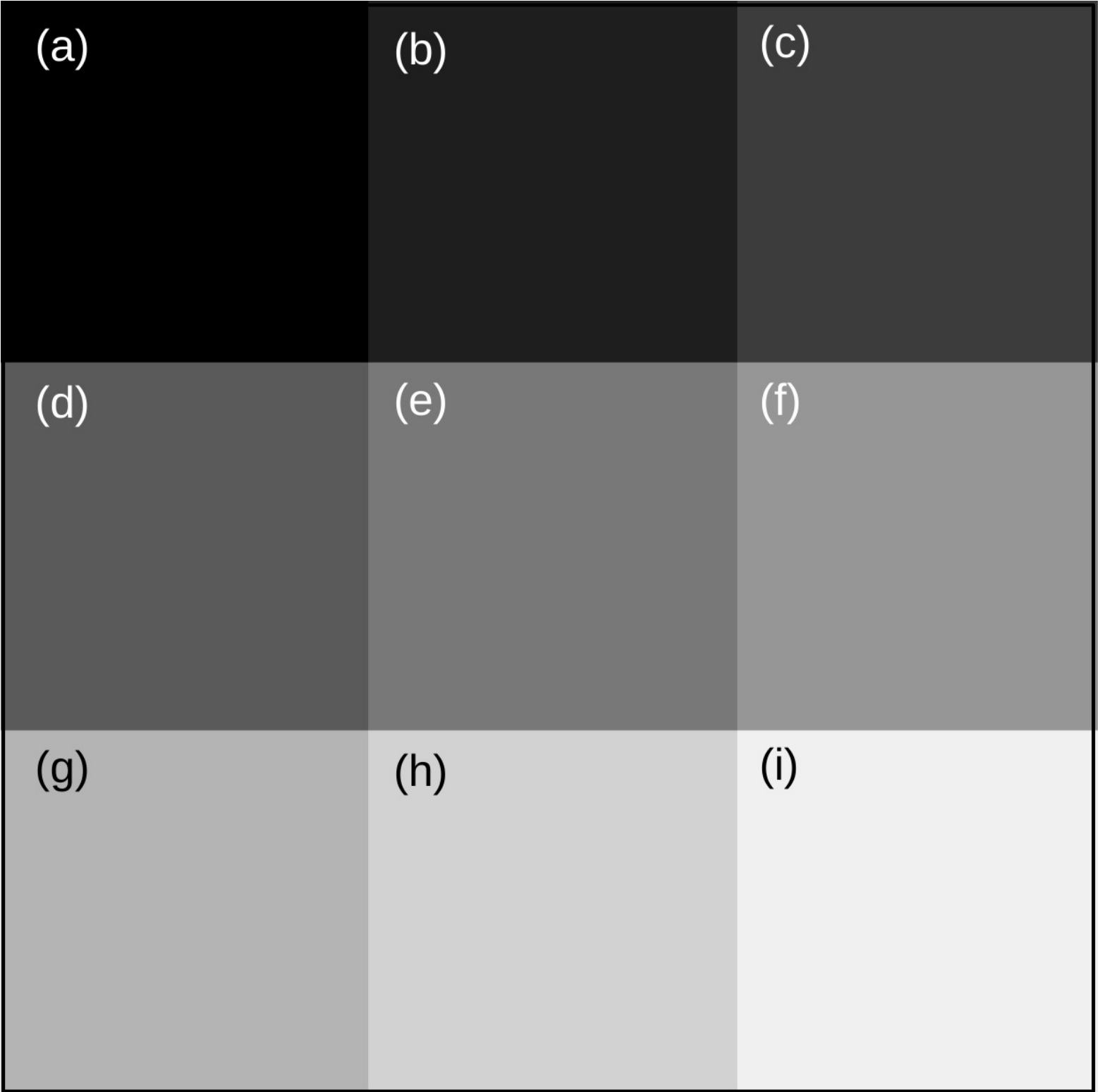}
\caption{Checkerboard gray level phase mask displayed on the SLM. The gray values shown in this phase mask are (a) g=0, (b) g=30, (c) g=60, (d) g=90, (e) g=120, (f) g=150, (g) g=180, (h) g=210 and (i) g=240. Each square in the checkerboard phase mask has different gray value which imparts differential phase shift to each beamlet in the incident structured beam on the SLM interferometer.}
\label{fig2}
\end{figure}

We setup a Michelson interferometer (Fig. \ref{fig1}) for our experiment, wherein one of the mirrors in one arm of the interferometer is replaced by the reflective SLM (LC-R 1080) from HOLOEYE Photonics AG. The reflective SLM has a pixel size of 8.1 $\mu$m. Note that similar effects may also be obtained using a Spatial Heterodyne Spectrometer (SHS) wherein both the mirrors are replaced by diffraction gratings. However, our experimental setup, where we use an SLM, provides almost realtime adaptive spectral tunability - which is much faster compared to a SHS setup. The detailed process of phase calibration of our reflective SLM is given in \cite{chandra2020rapid}. In brief, we use a broadband tunable light generated by a supercontinuum source (LEUKOS SAMBA) which is collimated by an optical fiber and then passes through a computer controlled electro-optic filter (LEUKOS BEBOP) having minimum and maximum adjustable bandwidths of 5 nm and 100 nm respectively. In our case, at first, we use the bypassed output from the electro-optic filter which passes broadband light in the wavelength range of 450-900 nm. The collimated light is expanded using a beam expander and then passes through a polarizer and a half-wave plate to select the linear polarization parallel to the director of liquid crystals in the SLM so as to operate the SLM in phase-only modulation configuration. The beam then passes through an aperture mask (schematic shown in Fig.~\ref{fig1}) having an array of $\mathrm{3\times 3}$ holes (each aperture having radius of about 2 mm) which generates a structured beam having an array of $\mathrm{3\times 3}$ beamlets which is incident on the SLM-based Michelson interferometer. The phase mask displayed on the SLM in our experiment is shown in Fig.~\ref{fig2} - it is basically a $\mathrm{3\times 3}$ checkerboard pattern consisting of different uniform gray level squares in the phase mask. The dimension of each square is set to 400 pixels utilizing the full width of our rectangular SLM display (1920 $\times$ 1200 pixels). Fig. 3(a) shows an array of $\mathrm{3\times 3}$ fringes recorded from the interferometer using a CCD camera. It is observed that the fringes in the individual beamlets show aberrations which are due to aberration in the experimental setup and the intrinsic phase fluctuation in the SLM display. This in turn would degrade the image quality of the field of view. An adaptive optics system needs to be used before the interferometer to measure and mitigate the aberration in the experimental setup. In addition, the SLM needs to be enclosed in a thermal oven and cooled to sub-zero temperatures to minimise the intrinsic phase fluctuation in the SLM and improve the imaging capability of the system. The spectrum measured using a fiber spectrometer for light directly
reflected from the SLM for different gray levels (60, 128, 200 and 240) is shown in Fig. 3(b). It should be noted that while these spectra display some intensity modulation, these are somewhat random, and certainly do not possess the periodicity as expected from fringes due to the SLM itself acting as a Fabry Perot cavity. On the other hand, the signal from the Michelson
interferometer (overplotted using a dashed line for gray value of 120) clearly shows
a fringe-like structure. The signal to noise of the fringes from the Micheslon
configuration are nearly 3.5 times higher than the oscillations observed because of
the intensity modulation due to the SLM response itself.

\subsection{Broadband hyperspectral imaging}

 We first demonstrate application of our novel method for broadband hyperspectral imaging. The broadband light in the wavelength range of 450-900 nm is incident on the SLM interferometer for this study. In the subsequent part of our work, we use a central wavelength of 637 nm and a bandwidth of 30 nm  by tuning the electro-optic filter adequately. This light is then incident on the SLM interferometer. We choose the central wavelength of 637 nm for potential spectroscopic applications of our method for astronomical applications in solar imaging, especially since the emission line at 637.4 nm of Fe X is an important constituent of the solar spectra \cite{swings1943edlen,habbal2011thermodynamics}. We display different gray level patterns on the SLM and measure the interferometer output by  using a fiber spectrometer (Ocean Optics USB4000 model having an order sorting filter according to specification). The resolution of the fiber based spectrometer is $\sim$0.26 nm. The checkerboard gray level pattern shown in Fig.~\ref{fig2} has gray values of 0, 30, 60, 90, 120, 150, 180, 210 and 240.
The gray value displayed on the SLM is mapped to the corresponding electric voltage at each individual pixel of the SLM using the SLM look-up table (ref. \cite{chandra2020rapid} for a sample of different look-up tables) and this voltage is applied between the transparent electrodes of the SLM display. This electric voltage induces birefringence $\Delta{n}$ across the parallel electrodes of the SLM display. The retardance generated in this process is proportional to the induced birefringence $\Delta{n}= n_{e}-n_o$, where $n_e$ is the extraordinary refraction coefficient and $n_o$ is the ordinary refraction coefficient of the liquid crystal in the SLM display. The phase difference imparted to light incident on the SLM is given by $\delta =2\pi \Delta{n} d/{\lambda}$ where $d$ is the distance between the parallel electrodes of the SLM and $\lambda$ is the wavelength of the light incident on the SLM. The phase difference imparted to each beamlet incident on different square gray value regions in the checkerboard phase mask is given by $\delta_i =2\pi \Delta{n_i} d/{\lambda}$ where i=1-9 (for the $\mathrm{3\times 3}$ checkerboard phase pattern), and $\Delta{n_i}$ is the induced birefringence due to the $i^{th}$ gray value in the checkerboard phase mask displayed on the SLM.

\begin{figure}
\centering\includegraphics[width=\linewidth]{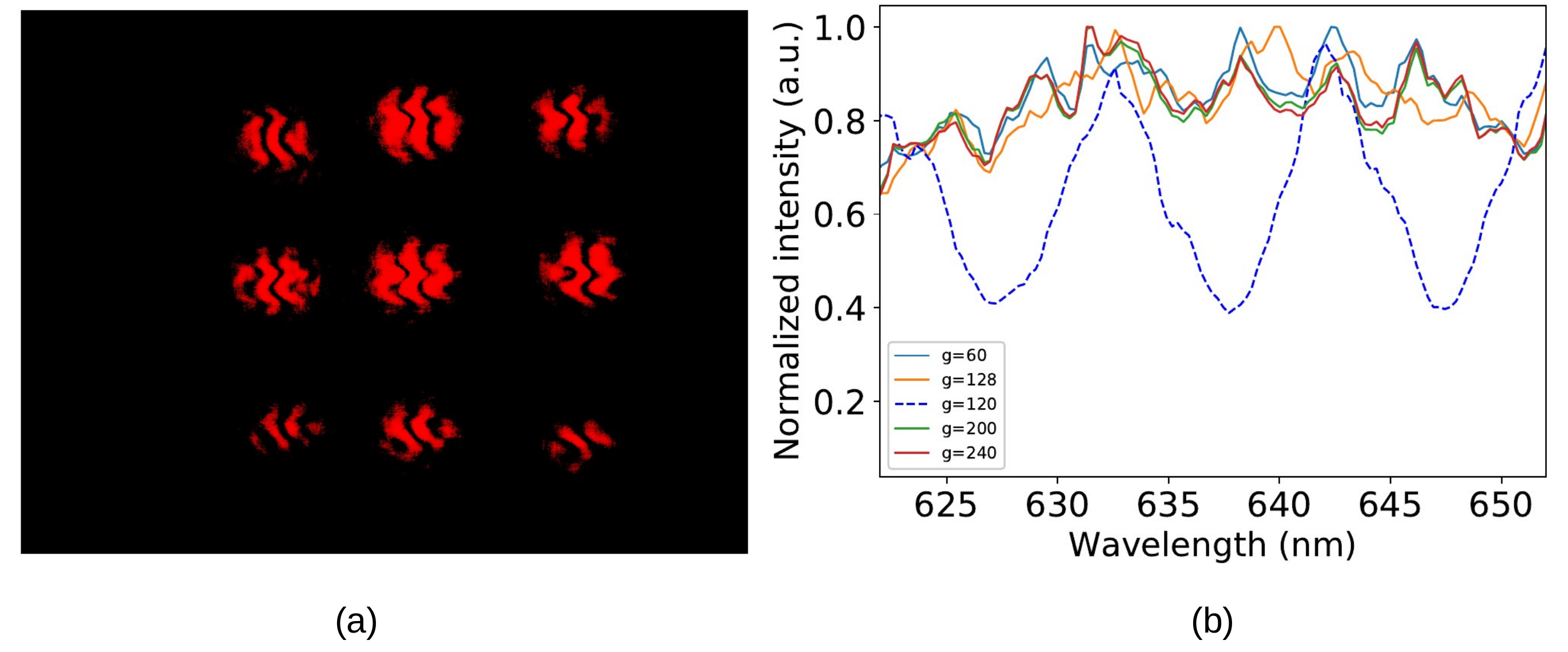}
\caption{(a) Array of 3 × 3 fringes recorded using CCD camera at the output port of the
Michelson interferometer. (b) Spectrum measured using a fiber spectrometer for light directly reflected from the SLM for different gray levels (60, 128, 200 and 240). The signal from the Michelson interferometer is overplotted using a dashed line for gray value of 120.}
\end{figure}

\begin{figure}
\centering\includegraphics[width=\linewidth]{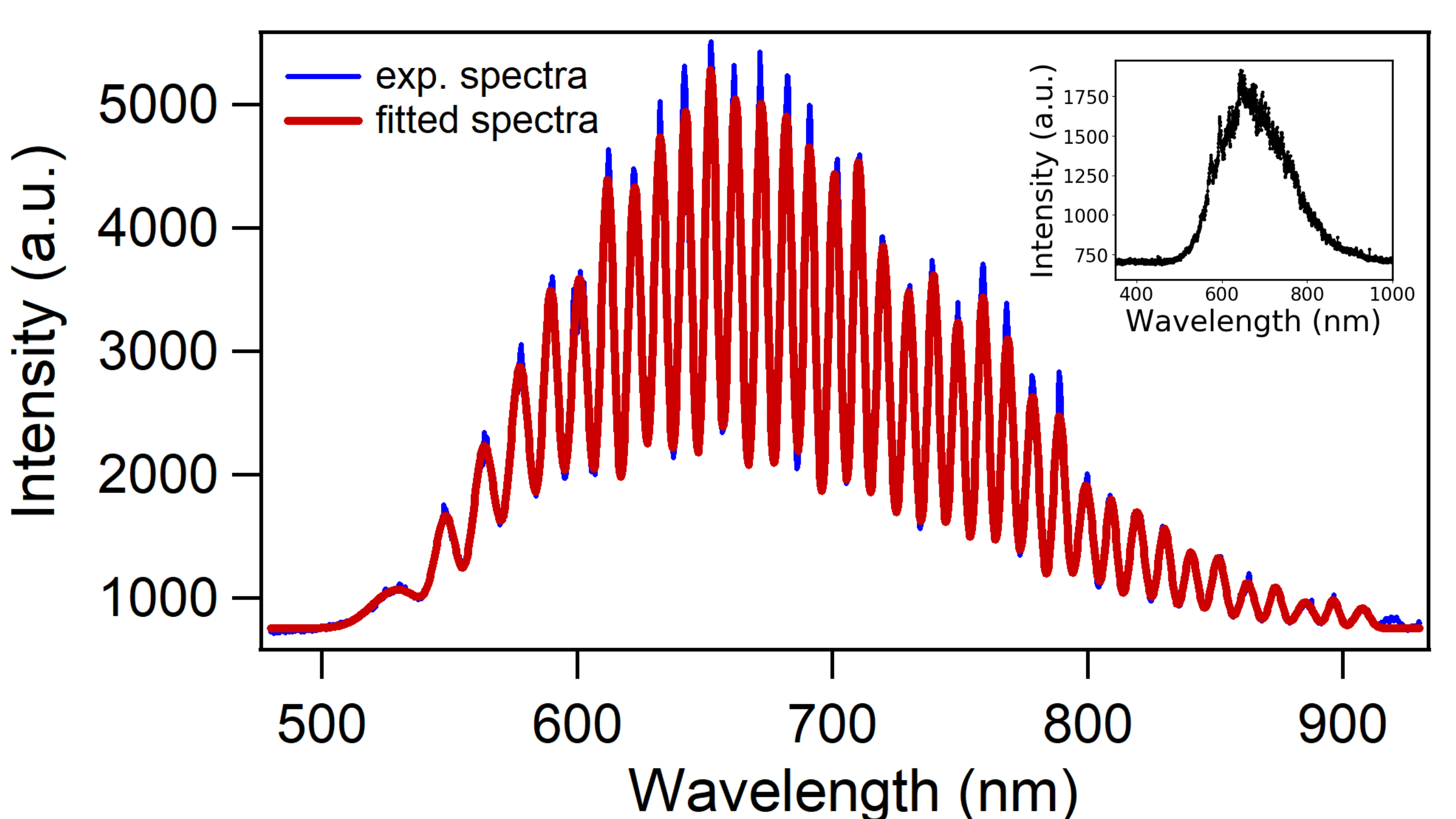}
\caption{Transmission spectrum measured from the output of the SLM interferometer using a fiber spectrometer for gray value of 120. The superposition of multiple Gaussians fitted to the spectrum is shown in red colour. The inset shows the spectrum of broadband illumination from the supercontinuum source measured using a fiber spectrometer.}
\label{fig4}
\end{figure}

In Fig. \ref{fig4} we show the recorded spectrum from the fiber spectrometer for broadband illumination from the supercontinuum source incident on the SLM interferometer. The inset in Fig. \ref{fig4} shows the source spectrum for the broadband light emanating from the supercontinuum source. We observe multiple well-resolved peaks in the spectrum obtained from the SLM interferometer having typical spectral resolution of about 3.8 nm (for peaks around 637 nm) which is limited by the light utilization efficiency (determined by the reflectivity of the SLM and also by the loss due to the pixelated structure of the SLM silicon backplane) of our reflective SLM (about 66\%). This paper concerns only about the fringe properties such as the free spectral range (FSR) and finesse which are discussed later rather than the dependence on the order number. The fiber spectrometer - besides having been provided with an inbuilt order sorting filter according to specifications - is sensitive to the intensity
modulation caused by the fringes, and displays an output in accordance with the fringe maximas and minimas along with their wavelength information. The estimated finesse of the interferometer ($F=\pi {(R_1 R_2)^{1/4}}/(1-(R_1 R_2)^{1/2})$) is about 14.4 using $R_1\sim$0.66 and $R_2\sim$0.975 where $R_1$ and $R_2$ are the reflectances of the SLM display and the silver mirror in the interferometer. However, it should be noted that the newer reflective SLMs having dielectric multilayer coatings have better reflectivity in the range of about 97\% which would enhance the spectral resolution of the resolved spectral peaks obtained from the SLM interferometer. The quoted reflectivity of about 97\% is for off the shelf SLMs and in addition the reflectivity of the SLM is dependent on wavelength. It might be possible to explore customized SLMs having specific coatings on the reflecting surface to enhance the reflectivity above 97\% in a relatively narrow wavelength range. This aspect needs to be explored in future using customized SLMs which might be useful to improve the spectral resolution to probe narrow spectral lines such as the 637.4 nm Fe X present in the solar spectrum. In addition, a pre-filter having narrow passband of about 1 nm can be used to improve the spectral resolution for the suggested application, so that higher arm length differences facilitated by the enhanced coherence length would result in lower FSRs and concomitantly higher fringe resolutions. 

Equally spaced multiple peaks are obtained in the transmission spectrum as they satisfy the resonance condition for a Michelson interferometer ($2\mu d= m \lambda$, where d is the path length difference between the two arms of the interferometer, $\mu$ is the refractive index of the medium, m in an integer and $\lambda$ is the wavelength of the incident light on the interferometer). The spacing between the peaks is about 11 nm which is the Free Spectral Range (FSR) of the SLM-based Michelson interferometer. The solid red curve in Fig.~\ref{fig4} shows the superposition of multiple Gaussians fitted to the spectrum. Fig. 5(a) shows overlapping spectra obtained for beamlets incident on two different gray levels of 0 and 180 present in the checkerboard phase mask displayed on the SLM. The relative shift of the resolved peaks in the two spectra for different gray level values of 0 and 180 is discernible. The top subplot in Fig. 5(a) shows the overlapping spectra zoomed in the wavelength range of 622-643 nm wherein relative shifts for the two spectral peaks around 625 nm and 636 nm at different gray values of 0 and 180 is detected. This demonstrates that the SLM-based Michelson interferometer can be tuned just by changing the gray value patterns displayed on the SLM, which eliminates the need for mechanical tuning of the interferometer. The bottom subplot in Fig. 5(a) shows the spectra from the SLM interferometer measured using the fiber spectrometer for two arm length differences of about 10 $\mu$m (red fringes) and 13 $\mu$m (blue fringes)(with the coherence length of the source being estimated between 15-20  $\mu$m) - which we set with the micrometer scale of resolution 1 $\mu$m attached to the mirror of the Michelson interferometer. The fiber based spectrometer is only for evaluating the adaptive hyperspectral imager and it will not be a part of the instrument. Equally spaced peaks are obtained in the transmission spectra as they satisfy the resonance condition for a Michelson interferometer. The spacing between spectral peaks for the 13 $\mu$m arm length difference is about 11 nm while that for the 10 $\mu$m arm length difference is about 17 nm. It is observed that the total number of peaks in a given transmission spectrum increases with increase in one of the arm lengths which is expected as the FSR is inversely proportional to the arm length difference of the interferometer. Thus, the FSR of the interferometer can be varied by changing the arm length of the interferometer. It is also observed that the bandwidth of the resonant peaks for smaller arm length difference is comparatively more (around 6 nm) than that for the larger arm length difference (around 4 nm) indicating poorer spectral resolution of the smaller interferometer. Fig. 5(b) shows the transmission spectra measured from the output of the Michelson interferometer wherein the SLM is replaced with a high reflectivity plane mirror. It is observed that the linewidth of the peaks around 637 nm narrows down to about 0.4 nm.\\

Fig.~\ref{fig6} shows the relative shift of the spectral peak around 636 nm by tuning the interferometer by varying the gray values displayed on the SLM display. The location of the spectral peak around 636 nm is obtained from fitting multiple Gaussians to the spectrum obtained for each gray level as demonstrated earlier in Fig.~\ref{fig4} for gray value of 120. Note that we have not implemented dispersion compensation in our experimental setup. This is because the wavelength response of the SLM varies by only 4.7\% over the
bandwidth of 30 nm we employ in the experiment. Similarly, the variation of the refractive index of air also changes only $3.72\times10^{-5}$ \%, so that both effects are smaller than the fringe drift due to phase fluctuations in the SLM. In addition, since we are not inferring any wavelengths using the interference fringes, but only demonstrating the capability of our technique, dispersion effects are not important here. The spectral peak around 636 nm seems to shift almost linearly with change in gray values from 0 to 255. The solid straight line shows the best linear fit to the relative phase excursions on varying the gray values displayed on the SLM, with the error
bars for each point being obtained by averaging the phase shift
for a set of 30 fringes for each value of gray level. From Fig. \ref{fig6}, the estimated minimum phase shift (from linear interpolation) imparted to the incident light beam is about 0.03 nm by changing the gray value by one unit. Our results demonstrate that the SLM-based Michelson interferometer can be used for broadband spectroscopic investigations of the desired target and does not involve any moving mechanical parts for tuning the interferometer. Also, as is evident from Fig. \ref{fig6}, the phase jitter is lowest between a gray value range of about 80 (100-180 in terms of absolute numbers). Note that this is also the region where the behaviour of our SLM is the most linear, as we had observed in our earlier work \cite{chandra2020rapid}. The typical phase jitter observed in this linear region (shown in the bottom panel in Fig. \ref{fig6}) is about 0.2-0.3 nm. However, this phase jitter/fluctuation is intrinsic to the SLM and can be partially mitigated to upto 80\% by putting the SLM in a thermal oven below zero degree temperatures as explored in an earlier work \cite{garcia2012flicker}. It would be challenging to probe the 637.4 nm line having width of about 0.1 nm given the phase jitter in our current SLM without cooling the SLM. However, this is not a major impediment of our possible suggestion to probe the 637.4 nm line using a SLM-based interferometer using newer SLMs which have low phase ripple (<1\%) compared to our SLM. Thus, we are somewhat restricted to using a little more than 6 bit useful phase depth out of the total available 8 bit operation (256 gray levels) of our SLM. This limitation can be mitigated by using SLMs having higher phase depth and increased total number of bits for phase modulation available in recent SLMs (16 bits). Assuming similar phase change of about 8 nm in SLMs having 1024 phase levels (10 bit), the spectral sampling would improve by a factor of about four (about 0.008 nm) compared to our spectral sampling using 256 phase levels (about 0.03 nm). This in turn would be very useful to probe narrow spectral lines such as the 637.4 nm Fe X line present in the solar spectrum. A broadband ($\sim$3.8 nm) imaging around the 637.4 nm line is feasible and for line resolved spectroscopy, the setup requires major modifications.

\begin{figure}
\centering\includegraphics[width=\linewidth]{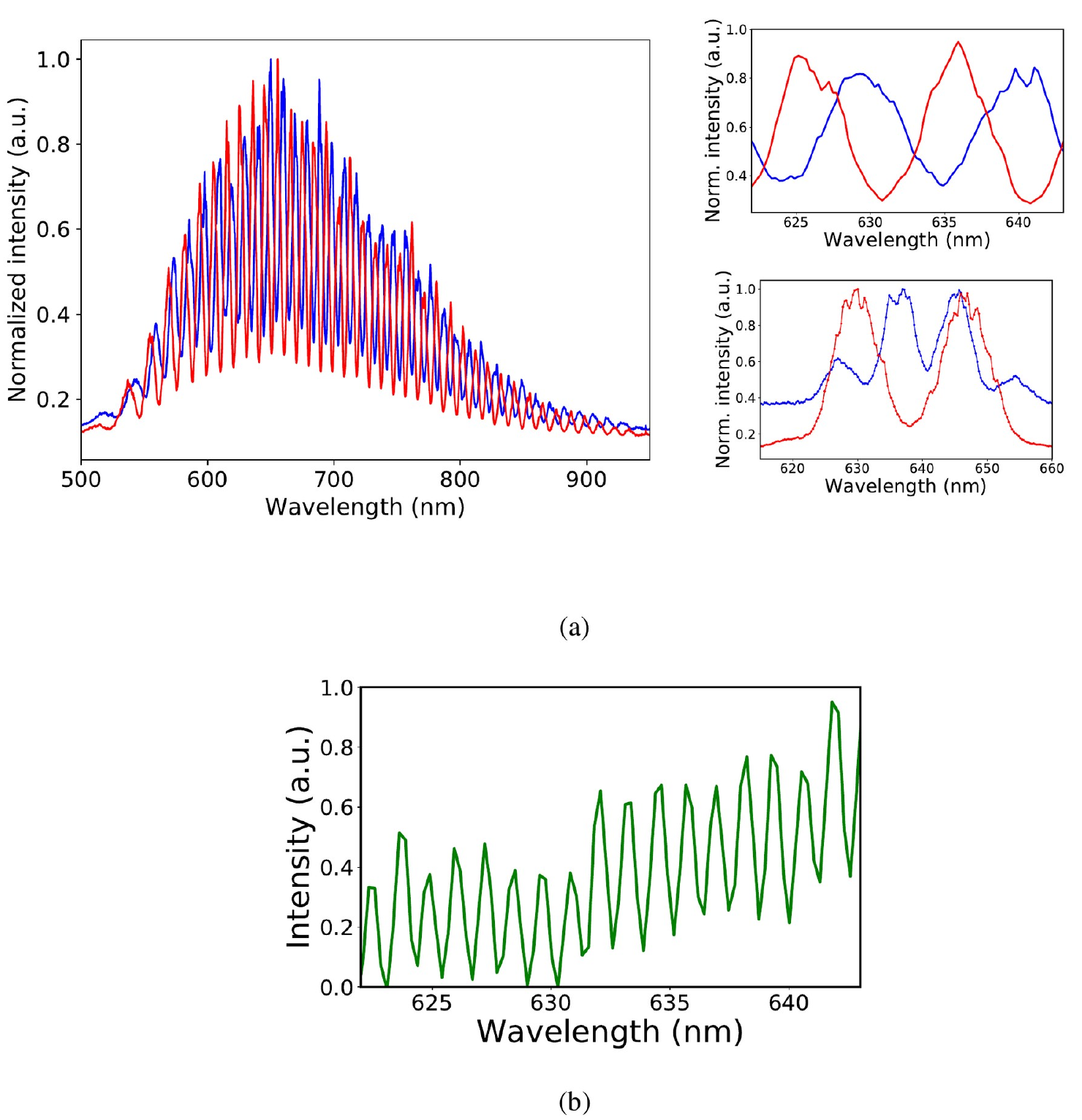}

\caption{(a) Transmission spectra measured from the output of the SLM interferometer using a fiber spectrometer for gray level values of 0 (red curve) and 180 (blue curve) shown in left panel. The top subplot shows the zoomed spectra in the wavelength range of 622-643 nm. Relative shifts between peaks in the spectrum with change in gray level is observed which demonstrates scanning of the interferometer by varying the gray level pattern displayed on the SLM. The bottom subplot shows transmission spectra measured from the output of the SLM interferometer using a fiber spectrometer for gray level value of 140 for two different arm length differences of about 10 $\mu$m  (red curve) and 13 $\mu$m (blue curve). (b) Transmission spectra measured from the output of the Michelson interferometer wherein the SLM is replaced with a high reflectivity plane mirror. The spectra is zoomed in the wavelength range of 622-643 nm.}
\end{figure}

\begin{figure}
\centering
\includegraphics[width=0.7\linewidth]{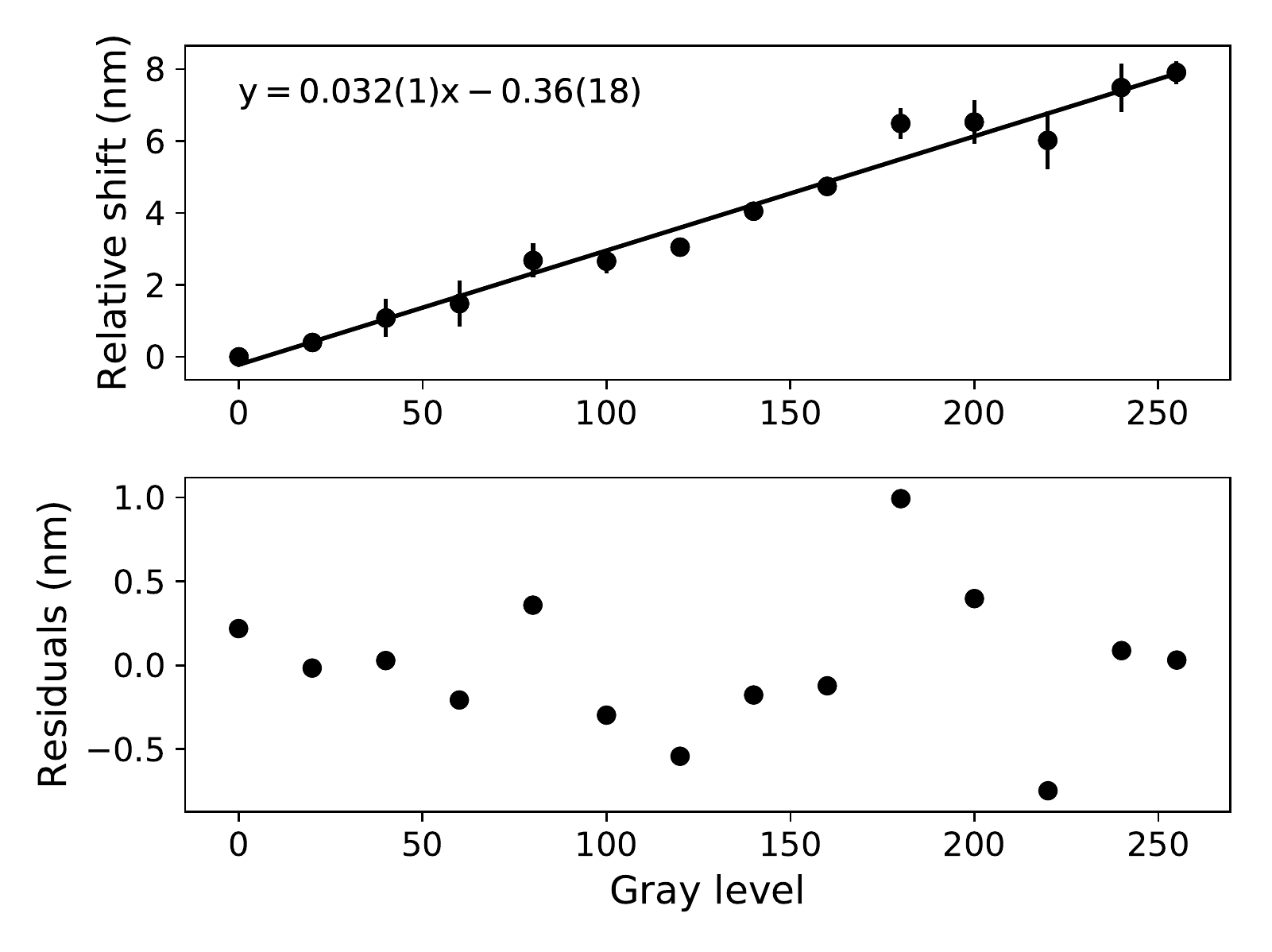}
\caption{Plot in the top panel showing relative shift in the spectral peak around 636 nm on scanning the SLM-based interferometer by varying the gray values displayed on the SLM. The best linear fit to the phase shift excursions vs gray level is shown by a straight line. Plot in the bottom panel showing residuals of the relative shift between the observations and the best linear fit.}
\label{fig6}
\end{figure}

\subsection{Adaptive hyperspectral imaging}

\begin{figure}
\centering\includegraphics[width=\linewidth]{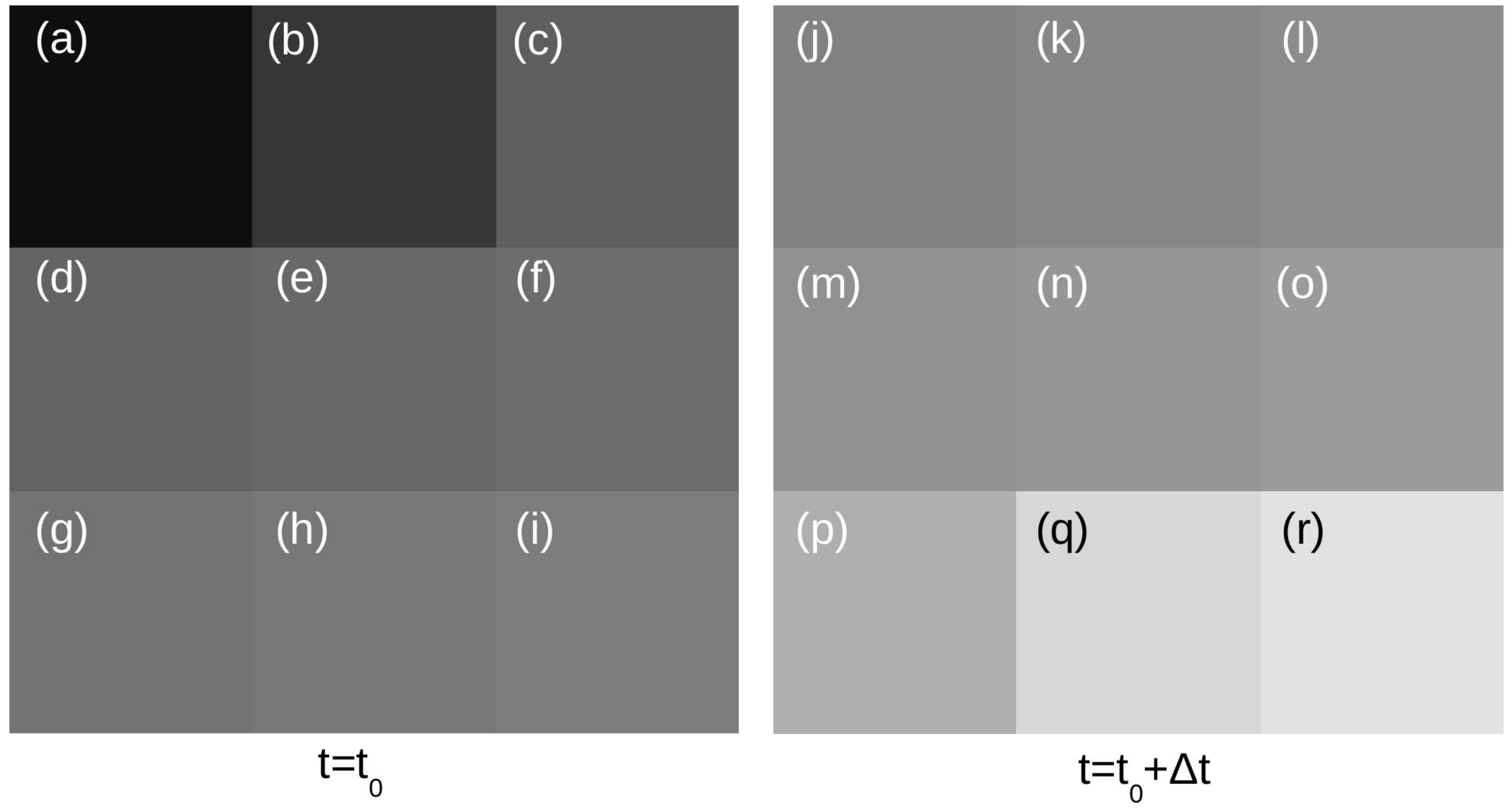}
\caption{Different checkerboard phase masks which can be used to obtain near contiguous spectral information using time multiplexing of the phase masks displayed on the SLM. The time interval between displaying the two phase masks is $\Delta$t where $\Delta$t $\geq$ response time of the SLM. The response time of our reflective SLM is about 18 ms. The gray values shown in the phase mask displayed at arbitrary reference time $t=t_0$ are (a) g=15, (b) g=55, (c) g=95, (d) g=100, (e) g=105, (f) g=110, (g) g=115, (h) g=120 and (i) g=125. The gray values shown in the phase mask displayed at time $t=t_0+\Delta t$ are (j) g=130, (k) g=135, (l) g=140, (m) g=145, (n) g=150, (o) g=155, (p) g=175, (q) g=215 and (r) g=255.}
\label{fig7}
\end{figure}

The phase mask shown in Fig.~\ref{fig2} is adaptive and can be updated with a new phase mask having different gray values in the checkerboard pattern in almost real-time (on time-scales of about 18 ms which is dependent on the response time of our SLM). This enables the recording of another set of near contiguous spectral information from the output of the SLM-based interferometer providing spectral tunability. Fig.~\ref{fig7} shows a set of two different checkerboard phase masks having different gray values which can be used to obtain near contiguous spectral information using time multiplexing of the phase masks. It should be noted that to achieve this kind of adaptive spectral tunability using spectral filters in a filter wheel makes the instrument more bulky and expensive. On the other hand tuning the interferometer by mechanical tuning in different steps requires comparatively more time and makes the instrument more prone to mechanical failures of moving components. Indeed, the tunability of the setup can be enhanced by using SLMs having lower response times which now reach about 1 kHz rate. The choice of the gray levels in the checkerboard phase mask can be staggered depending on requirements providing additional spectral tunability which is not available in fixed spectral filters. Our SLM has a modest power requirement of 24 V DC and can be useful for several applications having desirable modest power budget such as for space astronomy applications.\\ 

\subsection{Exploring long timescale phase stability of SLM}

\begin{figure}
\centering\includegraphics[width=0.7\linewidth]{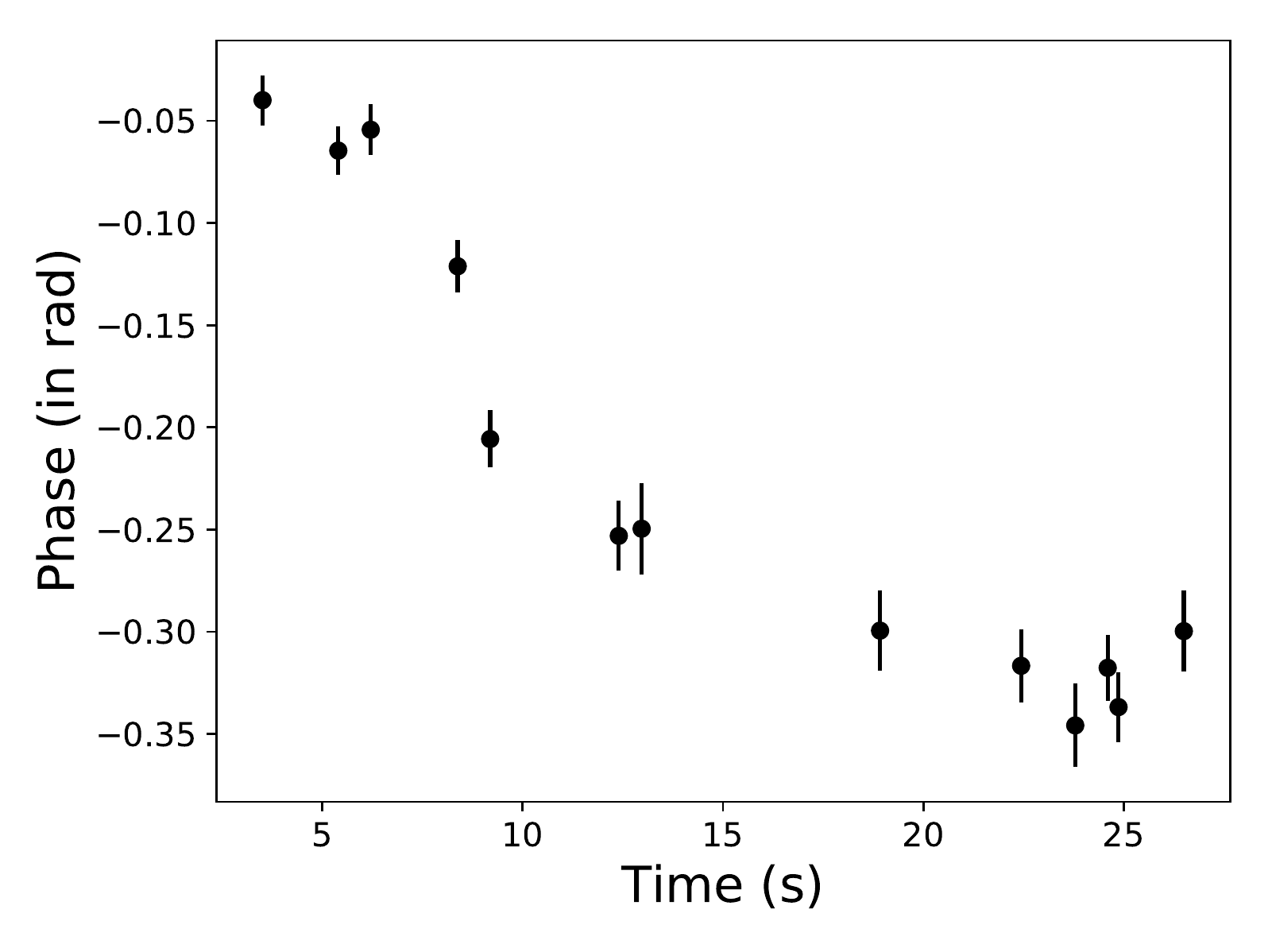}
\caption{Plot showing temporal phase variation in the SLM display using a SLM-based Michelson interferometer.}
\label{fig8}
\end{figure}

It has been shown that temporal phase flicker occurs in SLMs due to fluctuations in the orientation of liquid crystal molecules which are at the heart of the SLM display \cite{frumker2007phase,lizana2008time,garcia2012flicker,yang2019phase}. In an interesting study, \cite{garcia2012flicker} have shown that the phase flicker in SLMs can be reduced by about 80\% of the initial phase flicker by cooling the SLM display to $-8^{\circ}$C. We explore the temporal phase stability of our SLM by studying jitter in the interference fringes recorded from our Michelson interferometer. For this investigation we use a similar set-up as that shown in Fig. \ref{fig1}, and use an amplitude mask to generate a $\mathrm{2\times 2}$ multiplexed beam which falls on the SLM display. The phase mask displayed on the SLM is a uniform gray level mask having gray value of g=128 (we choose this value since this is the mean of the range where the SLM response is most linear). We record the interference fringes using a CCD camera setting integration time of 100 ms as we are interested in studying the phase stability of the SLM over time-scales of a few tens of seconds. We measure the relative position of fringes in each of the $\mathrm{2\times 2}$ array of fringes with respect to the leftmost fringe in each array. During our analysis we find that some of the fringes are of poor signal-to-noise ratio, so that we discard these from our analysis. Fig.~\ref{fig8} shows the relative phase of fringes with respect to the reference fringe. We find that the maximum phase flicker in our SLM is about 19\% over time-scales of about 26 s.
However, the phase flicker on shorter time-scales (about a few hundred ms) is relatively less (of the order of 0.1 radians). For bright astronomical sources such as the Sun, recording the spectrum using the SLM-based interferometer would require about a few hundreds of ms to build the spectrum without requirement for sequential scanning of the interferometer. We also observe that the phase flicker in each of the array of fringes is superposed on a downward phase drift trend which is intriguing.

\section{Discussions}
\subsection{Advantages of structured illumination in SLM-based adaptive interferometer}

The experimental setup shown in Fig.~ \ref{fig1} can achieve space-resolved spectroscopic investigation of the target whose spectral evolution occurs much slowly compared to the response time of the SLM, since the incident light is transformed into a structured beam using an amplitude mask. In this case, the array of beamlets will sample different regions in the field of view of the target and a series of uniform patterns having different gray level values needs to be displayed in sequence to capture spectral information for each beamlet. This method of obtaining spatial information of a given 2D field of view (FOV) has similar timescales (few milliseconds) involved in Fabry-Perot (FP) based spectroscopy. One of the advantages of SLM-based spectral scanning is availability of variable steps in tuning the interferometer which is usually not available in a conventional FP based setup. In addition, the power requirements to drive the SLM are a few tens of volts of DC compared to several hundreds of volts of AC power required to tune crystals based-FPs. One useful future addition to the setup might be to shift the beam across the amplitude mask using a tip-tilt mirror to obtain spectral signatures of different spatial regions of the target keeping the phase mask on the SLM unaltered, thus achieving additional spatial scanning of the target. We suggest another putative usage of our method wherein another SLM is used in tandem with the SLM-based Michelson interferometer. The incident beam may then be transformed into a structured beam using another SLM \cite{chandra2020efficient} instead of an amplitude mask. 
A holographic phase mask (HPM) can be displayed on an SLM to generate a $\mathrm{3\times 3}$ multiplexed beam from the input beam. The detailed method to generate this optimised phase mask using an iterative algorithm is discussed in \cite{chandra2020rapid, chandra2020efficient}. The biggest advantage of this suggested method is that the number of multiplexed beams can be changed in almost real-time on time-scales of about a few tens of ms. The checkerboard phase pattern displayed on the second SLM can be changed in sync with the first SLM so that each beamlet in the adaptive structured beam is incident on different gray value region displayed in the checkerboard phase  mask. Thus, both the optical element which produces the structured beam and that displaying the checkerboard phase pattern are adaptive in this case and might be useful in settings where the spectral signature of the target changes with time and the number of desired spectral channels can be configured in almost real time depending on requirements without modifying the experimental setup.\\ 

One can leverage the proposed adaptive setup to tune the spectral bands by increasing or decreasing the number of beamlets in the structured beam according to requirements and without modifying any hardware in the experiment setup. In addition, this proposed method does not involve any moving mechanical parts which is useful for potential spectroscopic applications using satellite based platforms. For implementation in the case of an astronomical source,  the  light from the source collected by a telescope would be incident on an amplitude mask which would generate a $\mathrm{N\times N}$ structured illumination. The structured beam should then be incident on a SLM-based Michelson/Fabry-Perot interferometer, wherein a $\mathrm{N\times N}$ checkerboard phase mask would be displayed on the reflective SLM such that each beamlet in the structured beam would fall on different regions in the checkerboard phase mask. This would impart a differential phase shift to each beamlet in the structured beam, thus generating an assemblage of the spectral information that is present in the incident beam in a single shot.

\section{Conclusions}

In summary, we have developed a novel hyperspectral imaging system that
takes advantage of structured illumination in an SLM interferometer. The design uses an adaptive checkerboard phase pattern displayed on the SLM to obtain single-shot near contiguous spectral information for each beamlet in the incident structured beam on the interferometer. The checkerboard phase pattern can be updated in near-real time on time-scales of about 18 ms (can be improved for SLMs having faster response) to capture another set of contiguous spectral bands in another shot depending on requirements. In addition, a pre-filter can be used to select the desired bandpass in the set-up providing additional spectral tunability of the setup. This design does not involve any moving parts and can be useful in settings requiring near real-time monitoring of spectral evolution of a scene or target for remote sensing applications. In addition, it can also be useful for single-shot multi-spectral investigations of static samples or targets thus reducing the time involved in spectral investigation. The design is compact, has low weight, near-real time tunability, low power requirements and does not involve moving mechanical parts making it ideally suited for multi-wavelength spectroscopic applications as required in solar spectroscopy and other applications in space astronomy including optical emission emanating from accretion discs around compact objects such as neutron stars and black holes, solar dynamic activity and optical follow-up of transient events such as $\gamma$-ray bursts. The improvements required to make this instrument from broadband spectral imaging to line resolved imaging include having customized coatings in SLM backplane for high reflectivity (>97\%), including a narrow band ($\sim$1 nm) pre-filter before the interferometer, enclosing the SLM in a thermal oven below zero degree temperatures and integrating an adaptive optics system in the setup. We are presently attempting validation of our technique with the sun itself as the source, and hope to report interesting results soon.

\section*{Funding}
Indian Institute of Science Education and Research Kolkata; Ministry of Education; Department of Science and Technology, Government of India.

\section*{Acknowledgements}
The Center of Excellence in Space Sciences India (CESSI) is funded by the Ministry of Education under the Frontier Areas of Science and Technology (FAST) scheme. ADC acknowledges support from the INSPIRE fellowship of the Department of Science and Technology, Govt. of India.

\section*{Disclosures}
The authors declare no conflicts of interest.

\section*{Data availability}
Data underlying the results presented in this paper are not publicly available at this time but may be obtained from the authors upon reasonable request.








\end{document}